\input harvmac

\lref\kakusp{Z. Kakushadze, Phys. Lett. B459 (1999) 497.}

\lref\kakuz{Z. Kakushadze, Nucl. Phys. B535 (1998) 311.}

\lref\bst{A. Buchel, G. Shiu and S.-H.H. Tye, {\it Anomaly
cancellation in orientifolds with quantised B flux}, hep-th/9907203.}

\lref\tensionless{M.J. Duff, R. Minasian and E. Witten, Nucl. Phys. 
B465 (1996) 413;\hfill\break
N. Seiberg and E. Witten, Nucl. Phys. B471 (1996) 121;\hfill\break
M. Berkooz, R.G. Leigh, J. Polchinski, J.H. Schwarz, N. Seiberg and 
E. Witten, Nucl. Phys. B475 (1996) 115.}

\lref\polch{J. Polchinski, Phys. Rev. Lett. 75 (1995) 4724.}

\lref\ads{I. Antoniadis, E. Dudas and A. Sagnotti, {\it Brane
supersymmetry breaking}, hep-th/9908023.}

\lref\gs{M.B. Green and J.H. Schwarz, Phys. Lett. B149 (1984) 117;
Nucl. Phys. B255 (1985) 93.}

\lref\dhvw{L. Dixon, J.A. Harvey, C. Vafa and E. Witten,
Nucl. Phys. B261 (1985) 678; \hfill\break 
Nucl. Phys. B274 (1986) 285.}

\lref\toroidal{M. Bianchi, G. Pradisi and A. Sagnotti,
Nucl. Phys. B376 (1992) 365.}

\lref\treshold{C. Bachas and C. Fabre, Nucl. Phys. B476 (1996)
418;\hfill\break 
I. Antoniadis, C. Bachas and E. Dudas, {\it Gauge couplings in 
four-dimensional type I string orbifolds}, hep-th/9906039.}

\lref\serone{C.A. Scrucca and M. Serone, {\it Anomalies and inflow on
D-branes and O-planes}, hep-th/9903145; {\it Anomaly cancellation in
K3 orientifolds}, hep-th/9907112.}

\lref\massimo{M. Bianchi, Nucl. Phys. B528 (1998) 73.}

\lref\witten{E. Witten, JHEP 02 (1998) 006.}

\lref{\ss}{A. Sen and S. Sethi, Nucl.Phys. B499 (1997) 45.} 

\lref\kst{Z. Kakushadze, G. Shiu and S.-H.H. Tye, 
Phys. Rev. D58 (1998) 086001.}

\lref\gp{E. Gimon and J. Polchinski, Phys.Rev. D54 (1996) 1667.}

\lref\cargese{A. Sagnotti, in Carg{\`e}se 87, Non-Perturbative Quantum Field
Theory, eds. G. Mack et al. (Pergamon Press, Oxford, 1988), p. 521.}

\lref\ps{G. Pradisi and A. Sagnotti, Phys. Lett. B216 (1989) 59.}

\lref\bs{M. Bianchi and A. Sagnotti, Phys. Lett. B247 (1990) 517}

\lref\twist{M. Bianchi and A. Sagnotti, Nucl. Phys. B361 (1991) 519.}

\lref\fps{D. Fioravanti, G. Pradisi and A. Sagnotti, Phys. Lett. 
B321 (1994) 349; \hfill\break
G. Pradisi, A. Sagnotti and Ya.S. Stanev, Phys. Lett. B345
(1995) 279; Phys. Lett. B356 (1995) 230; Phys. Lett. B381
(1996) 97.}

\lref\zero{A. Sagnotti, {\it Some properties of open string theories},
hep-th/9509080;\hfill\break 
A. Sagnotti, {\it Surprises in open string perturbation theory},
hep-th/9702093.}

\lref\zerob{C. Angelantonj, Phys. Lett. B444 (1998) 309;\hfill\break
R. Blumenhagen, A. Font and D. L{\"u}st, {\it Tachyon-free
orientifolds  of type 0B strings in various dimensions},
hep-th/9904069;\hfill\break
R. Blumenhagen and A. Kumar, {\it A note on orientifolds and dualities
of type 0B string theory}, hep-th/9906234.}

\lref\nonsusy{R. Blumenhagen and L. G{\"o}rlich, {\it Orientifolds of
non-supersymmetric asymmetric orbifolds}, hep-th/9812158;\hfill\break
C. Angelantonj, I. Antoniadis and K. F{\"o}rger, {\it
Non-supersymmetric type I strings with zero vacuum energy}, hep-th/9904092.}

\lref\gepner{C. Angelantonj, M. Bianchi, G. Pradisi, A. Sagnotti and Ya.S.
Stanev, Phys. Lett. B387 (1996) 743.}

\lref\ggsm{A. Sagnotti, Phys. Lett. B294 (1992) 196; \hfill\break
S. Ferrara, R. Minasian and A. Sagnotti, Nucl. Phys. B474 (1996)
323;\hfill\break
S. Ferrara. F. Riccioni and A. Sagnotti, Nucl. Phys. B519 (1998)
115; \hfill\break
F. Riccioni and A. Sagnotti, Phys. Lett. B436 (1998) 298.}

\lref\chiral{C. Angelantonj, M. Bianchi, G. Pradisi, A. Sagnotti and Ya.S. 
Stanev, Phys. Lett. B385
(1996) 96.}

\lref\adds{I. Antoniadis, E. Dudas and A. Sagnotti, Nucl. Phys. B544 (1999) 
469;\hfill\break
I. Antoniadis, G. D'Appollonio, E. Dudas and A. Sagnotti,
Nucl. Phys. B553 (1999) 133;\hfill\break
I. Antoniadis, G. D'Appollonio, E. Dudas and A. Sagnotti, {\it Open
descendants of $Z_2 \times Z_2$ freely-acting orbifolds}, hep-th/9907184.}

\lref\nsw{K.S. Narain, Phys. Lett. B169 (1986) 41;\hfill\break
K.S. Narain, M.H. Sarmadi and E. Witten, Nucl. Phys. B279 (1987) 369.}

\lref\pc{J. Polchinski and Y. Cai, Nucl. Phys. B296 (1988)
91;\hfill\break
G. Aldazabal, D. Badagnani, L.E. Ib{\'a}{\~n}ez and A.M. Uranga, {\it
Tadpole versus anomaly cancellation in $D=4$, $D=6$ compact IIB
orientifolds}, hep-th/9904071.} 

%%%%%%%%%%%%%%%%%%%%%%%%%%%%%%%%%%%%%%%%%%%%%%%%%%%%%%%%%%%%%%%%%%%%%%%%%%%%

\Title{\vbox{\rightline{\tt hep-th/9908064} \rightline{CPHT-S718.0599}
\rightline{LPTENS 99/27}}}
{\vbox{\centerline{Comments on Open-String Orbifolds }
\vskip .1in 
\centerline{with a Non-Vanishing $B_{ab}$}
}}

\centerline{Carlo Angelantonj\footnote{${}^\dagger$}{E-mail: 
{\tt angelant@cpht.polytechnique.fr}}}
\medskip
\centerline{\it Centre de Physique
Th{\'e}orique\footnote{${}^\ddagger$}{CNRS-UMR-7644}, 
{\'E}cole Polytechnique, F-91128 Palaiseau}
\centerline{\it and}
\centerline{\it Laboratoire de Physique 
Th{\'e}orique de l'{\'E}cole Normale Sup{\'e}rieure\footnote{${}^\star$}{
Unit{\'e} mixte de Reserche associ{\'e}e
au CNRS et {\'a} l'{\'E}cole Normale Sup{\'e}rieure.},}
\centerline{\it 24 rue Lhomond, F-75231 Paris Cedex 05}

\vskip 0.3in

\centerline{{\bf Abstract}}

\noindent
We study the effect of a non-vanishing flux for the NS-NS antisymmetric
tensor in open-string orbifolds. As 
in toroidal models, the total
dimension of the Chan-Paton gauge group is reduced proportionally to
the rank of $B_{ab}$, both on D9 and on D5-branes, while
the M{\"o}bius
amplitude involves some signs 
that, in the $Z_2$ orbifold, 
allow one to connect continuously ${\rm U} (n)$ groups to
${\rm Sp} (n) \otimes {\rm Sp} (n)$ groups on each set of
D-branes. In this case, non-universal couplings between twisted scalars
and gauge vectors arise, as demanded by the generalised
Green-Schwarz mechanism. We also comment on the role of the NS-NS
antisymmetric tensor in a recently proposed type I 
scenario, where supersymmetry is broken on the D-branes, while it is
preserved in the bulk.



\Date{8/99} %replace this line by \draft  for preliminary versions
             %or specify \draftmode at some point
%\draft

%%%%%%%%%%%%%%%%%%%%%%%%%%%%%%%%%%%%%%%%%%%%%%%%%%%%%%%%%%%%%%%%%%%%%%%

\newsec{Introduction}

Since the initial proposal to identify open strings as parameter space
orbifolds of closed oriented strings 
\cargese , much progress has
been made in our understanding of the rules that determine 
the construction of open-string vacua \refs{\bs,\ps,\fps}. 
In particular, the role played by modular
invariance for oriented closed strings is played by the tadpole
conditions and by a consistent interpretation of the various
amplitudes. Indeed, due to lack of modular
invariance, the Klein bottle, annulus and M{\"o}bius amplitudes develop
UV divergences that can be related to tadpoles
for unphysical Ramond-Ramond (R-R) fields \pc . Tadpole conditions,
then, ensure that the dangerous 
UV divergences are cured and, as result, the
total flux of the R-R fields vanishes consistently \polch.
 
A consistent interpretation means that all the amplitudes must provide
a coherent pattern of (space-time) massless and massive excitations in
the direct vacuum-channel, consistent with spin-statistics and, 
after $S$ or $P$ modular transformations, should describe a 
consistent tree-level dynamics of (projected) closed string states 
bouncing between holes and/or crosscaps. A
precise set of rules has been classified in \refs{\bs,\fps}, and can
be formally summarised as follows. Starting from a torus amplitude
$$
{\cal T} = \sum_{i,j} \chi_i N_{ij} \bar\chi_j
$$
that encodes the spectrum of the parent closed string theory, one
effectively realizes the world-sheet parity projection $\Omega$ by
adding the Klein bottle amplitude
$$
{\cal K} = {\textstyle{1\over 2}} \sum_i N_{ii} \chi_i \,.
$$
The restriction to the diagonal combinations $N_{ii}$ guarantees a
consistent particle interpretation, while an $S$-modular
transformation shows that the transverse-channel amplitude
$$
\tilde{\cal K} = \sum_i \Gamma_{i}^{2} \chi_i
$$
involves coefficients that are perfect squares and represent one-point
functions of closed fields in the presence of a crosscap.

Open unoriented strings correspond to the twisted sector of the world-sheet
orbifold and their relevant contributions to the one-loop 
partition function are encoded in the annulus amplitudes
$$
{\cal A} = {\textstyle{1\over 2}} \sum_{i,a,b} A^{i}_{ab} \chi_i n^a
n^b \,, \qquad 
\tilde{\cal A} = \sum_i B^{2}_{i} \chi_i \,,
$$
and in the M{\"o}bius amplitudes ($M^{i}_{a} = A^{i}_{aa}$ mod2)
$$
{\cal M} = {\textstyle{1\over 2}} \sum_{i,a} M^{i}_{a} \hat\chi_i n^a
\,, \qquad
\tilde{\cal M} = 2 \sum_i \Gamma_i B_i \hat\chi_i \,.
$$
Here $\tilde{\cal A}$ respects the rule of perfect squares with, now,
$B_i$ the one-point function of closed states in the presence of a
disk. Then, $\tilde{\cal M}$ is the geometric mean of $\tilde{\cal K}$
and $\tilde{\cal A}$, as expected since the M{\"o}bius strip has a
single boundary and a single crosscap. In the direct channel the
restriction to $M^{i}_{a} = A^{i}_{aa}$ mod2 ensures a consistent
particle interpretation in the open sector where the $n^a$'s represent
Chan-Paton multiplicities.

In this paper we shall apply these general rules to the case of
orbifold compactifications in the presence of a quantised NS-NS
antisymmetric tensor background\footnote{${}^{\P}$}{For a previous,
partial, description of orbifolds with a non-vanishing $B_{ab}$, 
without a complete construction of the partition functions, see
\kst.}. In toroidal
compactifications of the open strings \toroidal\ it was shown that,
although the corresponding modes are projected out by $\Omega$, a 
{\it quantised} $B_{ab}$ background is
still consistent with world-sheet parity. The main results in
\toroidal\ were that
the size the Chan-Paton gauge group is reduced in way that
depends on the
rank of $B_{ab}$, and that SO and Sp gauge groups may be
connected by (continuous) Wilson lines and sign ambiguities that 
enforce a proper normalisation of the M{\"o}bius amplitude. 
These results explained the key features of the gauge groups found
in the type-I compactifications studied in \refs{\bs,\twist}. 
Our main purpose here is to construct the full amplitudes (at a
generic, irrational, point in moduli space \refs{\ps,\gp}),
following the rules we
have previously summarised, and then to recover some
geometric insights  \witten\ added to the 
new phenomena discovered in \twist\ from
the constraints (and freedoms) imposed on the amplitudes.
We shall thus give an interpretation to the appearance of varying
numbers of tensor multiplets in the closed sector and we shall be able
to identify the signs present in ${\cal M}$ with discrete Wilson
lines.

This paper is organised as follows. In section 2 we review the
toroidal compactifications of the type I superstring in the 
presence of a non-vanishing $B_{ab}$. In section 3 we generalise this 
construction, discussing in detail the structure of the various
amplitudes for the $T^4/Z_2$ orbifold. In section 4 we give a
physical interpretation of the results presented in section 3, solving
the tadpole conditions and determining the massless excitations for
the various models. We also establish a connection between
discrete Wilson lines and the signs in ${\cal M}$, and,
finally, we comment
on the emergence of non-universal couplings between twisted scalars
and gauge vectors, consistently with the generalised Green-Schwarz
mechanism. In section 5 we generalise the construction of the open
descendants of the $Z$ orbifold \chiral , thus emphasising the role
played by the signs in the M{\"o}bius amplitude.
In section 6 we apply our results to a recently
proposed type I scenario \ads , where supersymmetry is broken on
branes but remains unbroken in the bulk. Section 7 contains our
conclusions.

%%%%%%%%%%%%%%%%%%%%%%%%%%%%%%%%%%%%%%%%%%%%%%%%%%%%%%%%%%%%%%%%%%%%%%%

\newsec{Review of generalised toroidal compactifications}

Let us review the compactification of open strings on a 
$d$-dimensional torus, studied for the first time in \toroidal .
On a generic $d$-dimensional lattice, the left and right momenta 
of the parent closed theory can be expressed as \nsw\
\eqn\momenta{
\eqalign{
p_a &= m_a + {\textstyle
{1\over \alpha'}} (g_{ab} - B_{ab} ) n^b\,,
\cr
\tilde p _a &= m_a - {\textstyle{1\over \alpha'}} (g_{ab} + B_{ab} ) n^b\,,
\cr}
}
where the metric $g_{ab}$ and the NS-NS antisymmetric tensor
$B_{ab}$ describe the size and the shape of the internal torus. 
It is evident that, at generic points in
moduli space, the parent theory is no more left-right symmetric.
Nonetheless, world-sheet symmetry still holds if one restricts
the attention to 
particular tori, for which effectively $p_a =
\tilde p_a$. As a result, the combination ${2 \over \alpha '} B_{ab}$ 
need be an integer, {\it i.e.} the antisymmetric tensor 
is quantised in appropriate units. This is consistent with the fact 
that, in the open descendants,
the fluctuations of the NS-NS two-tensor are projected out of the
spectrum, so that only the moduli of the internal
metric can be used to deform the lattice.

Once the left-right symmetry of the parent closed string is restored, one can 
proceed to construct the open descendants, starting, as usual, with the
Klein bottle amplitudes
$$
{\cal K}^{({\rm tor})} = {\textstyle{1\over 2}}
(V_8 - S_8 ) (q^2 ) \sum_{m} {q^{{\alpha ' \over
2} m^{{\rm T}} g^{-1} m} \over \eta^d (q^2)} 
$$
and
$$
\tilde{\cal K}^{({\rm tor})} = {2^5\over 2} 
\, \sqrt{{\rm det} (g/\alpha ')} \,\, 
(V_8 - S_8 ) ({\rm i} \ell )
\sum_{n} {({\rm e}^{-2 \pi \ell})^{{1\over \alpha '} n^{{\rm T}} g n} \over
\eta^d ({\rm i} \ell )} \,.
$$
Here and in the following, 
a tilde denotes amplitudes in the transverse channel, and ``vertical'' 
and ``horizontal'' proper times are related by an $S$ 
modular transformation for the Klein bottle 
and annulus amplitudes and by a $P=T^{{1\over 2}}ST^2 S T^{{1\over 2}}$ 
transformation for the M{\"o}bius amplitude. 

The construction of the open-string sector presents some interesting 
new features.
Following \refs{\cargese,\bs}, the transverse-channel annulus amplitude 
involves
only characters that fuse into the identity with their anti-holomorphic
partners in the closed-string GSO. In our case, this translates in the
restriction to states with $p_a = - \tilde p_a$. Given the quantisation
condition on the $B$-field, only the winding states satisfying
\eqn\quant{
{\textstyle{2 \over \alpha '}} B_{ab} n^b = 2 m_a 
}
can contribute to the transverse-channel annulus amplitude.
Inserting a projector that effectively realizes \quant , one gets
$$
\tilde{\cal A} ^{({\rm tor})} = {2^{r-d-5} \over 2}
\, \sqrt{{\rm det} (g/\alpha ') }
\,\,\, N^2  \, (V_8 - S_8 )({\rm i}\ell)
\sum_{\epsilon =0,1}
\sum_{n} {({\rm e}^{-2 \pi\ell})^{{1\over 4 \alpha '} n^{{\rm T}} g n}
\,{\rm e}^{{2{\rm i} \pi \over \alpha '} n^{{\rm T}} B \epsilon} \over \eta^d
({\rm i}\ell )}
$$
and
$$
{\cal A}^{({\rm tor})} = {2^{r-d} \over 2}\,\,\, 
N^2 \, (V_8 - S_8 ) (\sqrt{q}) 
\sum_{\epsilon =0,1} \sum_{m}  {q^{{\alpha ' \over 2} (m + 
{1\over \alpha '} B
\epsilon) ^{{\rm T}} g^{-1} ( m + {1\over \alpha '} B \epsilon )} 
\over \eta^d (\sqrt{q})}
\,,
$$
where $r={\rm rank}(B)$, and the normalisation in ${\cal A}$ 
guarantees that the
massless vector has the right multiplicity. 

The M{\"o}bius amplitudes
$$
\eqalign{
\tilde{\cal M}^{({\rm tor})} = - {2 \times 2^{(r-d)/2} \over 2} 
\, \sqrt{{\rm det} (g/\alpha ')} \,&\,\, 
N \,\, (\hat V _8 - \hat S _8 ) ({\rm i}\ell+
{\textstyle{1\over 2}}) \times 
\cr
& \times 
\sum_{\epsilon =0,1} \sum_{n} {({\rm e}^{-2\pi\ell})^{{1\over \alpha '}
n^{{\rm T}} g n} \,\, {\rm e}^{{2{\rm i} \pi \over \alpha '} 
n^{{\rm T}} B \epsilon}
\,\, \gamma_\epsilon \over \hat\eta^d ( {\rm i}\ell {\textstyle{1\over 2}})}
\cr}
$$
and
$$
{\cal M}^{({\rm tor})} = - {2^{(r-d)/2}\over 2} \,\,\, 
N\,\, (\hat V _8 - \hat S _8 ) (-\sqrt{q}) 
\sum_{\epsilon =0,1} \sum_{m} {q^{{\alpha ' \over 2} (m + 
{1\over \alpha '} B
\epsilon )^{{\rm T}} g^{-1} (m + {1\over \alpha '} B \epsilon )} 
\, \, \gamma_\epsilon \over
\hat\eta^d (-\sqrt{q} ) }
$$
complete the $\Omega$-projection in the open-string sector. As is
\bs , in order to
compensate the fixed real part of the modulus of the doubly-covering torus,
we have introduced real ``hatted'' characters. 
Notice that the M{\"o}bius amplitudes involve crucial signs 
$\gamma_\epsilon$ that enforce a correct normalisation of the
$B$-field projector. 

The last step in the construction of open descendants 
consists in imposing the cancellation of tadpoles of unphysical
massless states that flow in the transverse
channel. This gives the well known result
$$
N = 2^{5-r/2} \,,
$$
{\it i.e.} a non-trivial quantised background for the NS-NS
antisymmetric tensor of rank $r$ 
reduces the rank of the Chan-Paton gauge group by a factor
$2^{r/2}$. In geometrical terms, the identification $B={1\over 2} 
w_2$, with $w_2$ the generalised second Stieffel-Whitney
class, suggests to interpret the presence of a quantised $B_{ab}$ as an
obstruction to define a vector structure on the 
compactification torus \refs{{\ss},\massimo,\witten}.

Before closing this section, let us comment on the role of the signs
$\gamma_\epsilon$ that must be introduced in the M{\"o}bius amplitude.
To this end, it is convenient to analyse in some detail the case of a
two-torus. The metric $g_{ab}$ and the antisymmetric tensor can be
conveniently parametrised as
$$
g = {\alpha ' Y_2 \over X_2} \left( \matrix{ 1 & X_1 \cr X_1 & |X|^2
\cr}
\right) \ , \qquad B = {\alpha' \over 2} \left( \matrix{0 & Y_1 \cr -
Y_1 & 0 \cr} \right) \ ,
$$
where $X$ and $Y$ define the complex and K{\"a}hler structures,
respectively.

The corresponding annulus amplitudes are
$$
\tilde{\cal A} = {2^{-5} \over 2} \, Y_2 \, N^2
(V_8 - S_8 ) ({\rm i} \ell) \sum_{n_1 ,n_2} {W_{n_1 ,n_2} \left[ 1 +
(-1)^{n_1} + (-1)^{n_2} + (-1)^{n_1 + n_2} \right] \over \eta^2 ({\rm
i} \ell)} \,,
$$
$$
{\cal A} = {\textstyle{1\over 2}} N^2 (V_8 - S_8)
(\sqrt{q}) \, {P_{0,0} + P_{0,1} + P_{1,0} + P_{1,1} \over \eta^2
(\sqrt{q})} \,,
$$
and, in a similar fashion, the M{\"o}bius amplitudes are
$$
\eqalign{
\tilde{\cal M} =& - {\textstyle{2\over 2}} \, Y_2 \, N
\, (\hat V_8 - \hat S_8) ({\rm i} \ell + {\textstyle{1\over 2}} ) \,
\sum_{n_1 , n_2} {W_{n_1 , n_2} \over \hat\eta^2 ({\rm i} \ell +
{\textstyle{1\over 2}} )} 
\cr
&\times \left[ \gamma_{0,0} + (-1)^{n_1}
\gamma_{0,1} + (-1)^{n_2} \gamma_{1,0} + (-1)^{n_1 +n_2} \gamma_{1,1}
\right] \,,
\cr}
$$
$$
{\cal M} = -{\textstyle{1\over 2}} \, N \, (\hat V _8
- \hat S_8) (-\sqrt{q}) \, {\left[ \gamma_{0,0} P_{0,0} + \gamma_{0,1}
P_{0,1} + \gamma_{1,0} P_{1,0} + \gamma_{1,1} P_{1,1} \right] \over
\hat\eta^2 (-\sqrt{q})} \,.
$$
In these expressions we have introduced a compact notation for the
winding and momentum sums:
$$
\eqalign{
W_{n_1 , n_2 } &= ({\rm e}^{-2 \pi \ell})^{{Y_2 \over 4 X_2} 
[(n_1 + X_1 n_2)^2 + n^{2}_{2}
X^{2}_{2} ]} \,,
\cr
P_{\epsilon_1 , \epsilon_2} &= \sum_{m_1 , m_2} q^{{1\over 2 X_2 Y_2}
[(m_1 + {\epsilon_2 \over 2} - X_1 (m_2 - {\epsilon_1 \over 2}))^2 +
(m_2 - {\epsilon_1 \over 2})^2 X^{2}_{2} ]} \,.
\cr}
$$
Expanding in powers of $q$, one can easily find the contribution of
each amplitude to the tadpole of the R-R 10-form:
$$
\eqalign{
\tilde{\cal K} &\sim 2^5 \,,
\cr 
\tilde{\cal A} &\sim 2^{-5} \times N ^2 \times 4 \,,
\cr
\tilde{\cal M} &\sim -2 \times N \times (
\gamma_{0,0} + \gamma_{0,1} + \gamma_{1,0} + \gamma_{1,1} ) \,.
\cr}
$$
It is then evident that, in order to solve the tadpole condition, one
of the four $\gamma_\epsilon$ has to equal minus one, while the
three others have to equal plus one. Among the four possible choices,
only two lead to different results and differ in the sign of
$\gamma_{0,0}$. If $\gamma_{0,0}=+1$, the massless vector in the
direct channel belongs to the adjoint of an ${\rm SO} (16)$ gauge
group, while if $\gamma_{0,0} =-1$ it belongs to a symplectic group. 

These signs $\gamma_\epsilon$ are related to the existence of
different kinds of orientifold planes with opposite R-R charge
\witten . As we shall see in the next sections, a careful construction
of the partition functions reveals neatly their role, and shows that
they are also responsible for similar
``group transitions'' in orbifold compactifications.

%%%%%%%%%%%%%%%%%%%%%%%%%%%%%%%%%%%%%%%%%%%%%%%%%%%%%%%%%%%%%%%%%%%%%%%%%

\newsec{The $T^4 /Z_2$ orbifold with a non-vanishing $B_{ab}$}

Let us now turn to the compactification on the irrational $T^4 /Z_2$ orbifold, 
in the presence of a non-vanishing background for the NS-NS
antisymmetric tensor.  

To this end let us introduce the combinations \bs
$$
\eqalign{
Q_O &= V_4 O_4 - C_4 C_4 \,,
\cr
Q_V &= O_4 V_4 - S_4 S_4 \,,
\cr}
\qquad
\eqalign{
Q_S &= O_4 C_4 - S_4 O_4 \,,
\cr
Q_C &= V_4 S_4 - C_4 V_4 \,,
\cr}
$$
of level one ${\rm SO} (4)$ characters, 
that represent the contributions of the world-sheet fermions
to the partition function. They are eigenstates of the $Z_2$ generator, with 
eigenvalues $\pm 1$ for $(Q_O, Q_S)$ and $(Q_V ,Q_C)$, respectively. 
Including the contributions of the internal bosons, the 
partition function for the parent type IIB superstring then reads:
\eqn\torus{
\eqalign{
{\cal T} =& {1\over 2} \Biggl[ |Q_O + Q_V |^2 
\sum_{m\,,\, n} {
q^{{\alpha ' \over 4}  p g^{-1} p} \,
\bar q^{{\alpha ' \over 4} \tilde p g^{-1} \tilde p} 
\over |\eta^4|^2} 
+ |Q_O - Q_V |^2 \, \left|{\vartheta^{2}_{3} \vartheta^{2}_{4} 
\over \eta^4} \right|^2 +
\cr
& \quad + |Q_S + Q_C |^2 \, \left| {\vartheta^{2}_{2}
\vartheta^{2}_{3} 
\over  \eta^4}\right|^2
+ |Q_S - Q_C |^2 \, \left| {\vartheta^{2}_{2} \vartheta^{2}_{4} 
\over \eta^4}\right|^2
\Biggr]\,,
\cr}
}
where we have omitted the integration measure and the contributions
of non-compact bosonic coordinates.
In order to extract the massless spectrum and determine the correct 
expressions for the direct-channel Klein bottle amplitude and the 
transverse-channel annulus amplitude, one can extract the contributions
at the origin 
from the lattice sums and rewrite the torus amplitude in terms
of ``generalised'' characters: 
\eqn\torusmz{
\eqalign{
{\cal T}_{0} 
\sim & |Q_O \phi_O + Q_V \phi_V |^2 + |Q_O \phi_V + Q_V \phi_O |^2 +
\cr
& + 16 \left( |Q_S \phi_S + Q_C \phi_C|^2 + |Q_S \phi_C + Q_C \phi_S 
|^2 
\right) 
\,,
\cr}
} 
where we have introduced the combinations 
$$
\eqalign{
\phi_O &= {\textstyle{1\over 2}} \left( {1\over \eta^4} + {\vartheta_{3}^{2}
\vartheta_{4}^{2} \over \eta^4} \right) \,,
\cr
\phi_V &= {\textstyle{1\over 2}} \left( {1\over \eta^4} - {\vartheta_{3}^{2}
\vartheta_{4}^{2} \over \eta^4} \right) \,,
\cr}
\qquad
\eqalign{
\phi_S &=  
{\textstyle{1\over 8}} \left( {\vartheta_{2}^{2} \vartheta_{3}^{2} 
\over \eta^4} +  {\vartheta_{2}^{2}\vartheta_{4}^{2} \over 
\eta^4} \right) \,,
\cr
\phi_C &= {\textstyle{1 \over 8}} \left(
{\vartheta_{2}^{2} \vartheta_{3}^{2}
\over \eta^4} - {\vartheta_{2}^{2}\vartheta_{4}^{2} \over \eta^4} \right) \,.
\cr}
$$
The numerical factor in \torusmz\ counts the number of points 
left fixed by the  action of the orbifold generator, and thus gives 
multiplicities to states in the twisted sector.  
For a $Z_2$ orbifold it is equal to 16, as one expects from the
Lefschetz theorem \dhvw.

Expanding the ``generalised'' characters to the leading (massless) 
order in $q$, one gets
$$
\eqalign{
Q_O \phi_O + Q_V \phi_V &\sim V_4 - 2 C_4 \,,
\cr
Q_O \phi_V + Q_V \phi_O &\sim 4 O_4 - 2 S_4 \,,
\cr}
\qquad
\eqalign{
Q_S \phi_S + Q_C \phi_C &\sim 2 O_4 - S_4 \,,
\cr
Q_S \phi_C + Q_C \phi_S &\sim {\rm massive} \,,
\cr}
$$
from which one can easily read the spectrum of the $Z_2$ orbifold of the type 
IIB superstring.
The untwisted sector comprises  the ${\cal N}=(2,0)$ supergravity 
multiplet and five tensor multiplets, while the twisted sector comprises 16 
more tensor multiplets, one from each fixed point. 
Thus, the full spectrum is anomaly-free and 
corresponds to the compactification on a smooth K3.

It has been shown in \refs{\kst , \witten} that, in the presence of a
quantised $B_{ab}$, not all fixed points have the same $\Omega$-eigenvalue.
For instance, for the 2d toroidal compactification of type ${\rm I}'$ in the
presence of a non-trivial background for the 
antisymmetric tensor, three of the four fixed points have a
positive eigenvalue, whereas the fourth one has a negative
eigenvalue\footnote{${}^\dagger$}{The $\Omega$-eigenvalues of the 
fixed points are strictly related to the signs $\gamma_\epsilon$ 
that we have discussed in the previous section.}. 
The generalisation to the case of a $T^4/Z_2$ orbifold is
straightforward and, for a generic $B_{ab}$ with rank $r$, the number
of positive and negative eigenvalues is
\eqn\npm{
n_\pm = 2^3 (1  \pm 2^{-r/2} ) \,.
}
As we shall see in a moment, this is of crucial importance in order to
obtain a consistent transverse Klein bottle amplitude.

In orbifold compactifications, states coming from twisted sectors are
localised on (invariant combinations of) fixed points. As a result, in
computing the Klein-bottle amplitude
$$
{\cal K} = {\textstyle{1\over 2}} \, {\rm tr}_{{\cal H}_c}\, 
\Omega \, q^{L_0} \,\bar q ^{\bar L_0} 
$$
one has to combine the action of world-sheet parity on closed-string
states with the action of world-sheet parity on fixed points. Thus,
taking into account \npm, from \torusmz, one can deduce the following
expression for the Klein-bottle amplitude restricted to the low-lying 
excitations:
$$
\eqalign{
{\cal K}^{(r)}_{0} \sim & {\textstyle{1\over 2}} \Bigl\{
(Q_O \phi_O + Q_V \phi_V ) + (Q_O \phi_V + Q_V \phi_O ) +
\cr
& + (n_+ - n_-) \left[ (Q_S \phi_S + Q_C \phi_C ) + (Q_S \phi_C + Q_C
\phi_S ) \right] \Bigr\} \,,
\cr}
$$
from which one can immediately read the spectrum of massless
excitations. Beside the ${\cal N} =(1,0)$ gravitational multiplet
coupled to the universal tensor multiplet and four hypermultiplets
from the untwisted sector, one gets $n_+$ hypermultiplets and $n_-$
tensor multiplets from the twisted sector, as a result of the different
behaviour of the $Z_2$ fixed points under $\Omega$, {\it i.e.} 
of the different charge of the orientifold planes.

The full Klein-bottle amplitude can be computed adding the
contributions from massive states filling a sublattice of the original
$T^4$. Due to the $Z_2$ orbifolding, both momentum and winding states
contribute to the amplitude. Actually, the latter have to satisfy
the constraint \quant. As a result, one has to introduce a projector
in the sum over winding states, so that the full amplitude reads
$$
\eqalign{
{\cal K}^{(r)} =& {\textstyle{1\over 4}} (Q_O + Q_V)(q^2 ) \, 
\left[ \sum_m {q^{{\alpha ' \over 2} m^{{\rm T}} g^{-1} m } \over
\eta ^4 (q^2)} + 2^{-4} \sum_{\epsilon =0,1} \sum_n 
{q^{{1\over 2\alpha '} n^{{\rm T}} g n} \, {\rm e}^{{2 {\rm i} \pi
\over \alpha '} n^{{\rm T}} B \epsilon} \over \eta ^4 (q^2)} \right]
\cr
&+ {2^{(4-r)/2} \over 2} \, (Q_S + Q_C) (q^2 ) \left( {\vartheta ^{2}_{2}
\vartheta ^{2}_{3} \over \eta ^4} \right) (q^2 ) \,.
\cr}
$$
The coefficient in front of the winding sum ensures the correct
normalisation of the graviton. After an $S$ modular transformation, one
gets the following expression for the transverse channel Klein-bottle
amplitude:
$$
\eqalign{
\tilde{\cal K} =& {2^5 \over 4} (Q_O + Q_V ) ({\rm i} \ell )
\Biggl[ {\rm Vol} \, \sum_n {({\rm e}^{-2 \pi\ell})^{{1\over \alpha '}
n^{{\rm T}} g n} \over \eta ^4 ({\rm i}\ell)}  +
\cr
& \qquad\quad +
{2^{-4}\over {\rm Vol}} \, \sum_{\epsilon=0,1} \sum_m { ({\rm e}^{-2\pi
\ell})^ {\alpha ' (m + {1\over \alpha '} B \epsilon )^{{\rm T}} g^{-1}
(m + {1\over \alpha '} B\epsilon)} \over \eta ^4 ({\rm i}\ell )}
\Biggr] +
\cr
&+ {2^{5-r/2}\over 2} (Q_O - Q_V) ({\rm i}\ell ) \left( {\vartheta
^{2}_{3} \vartheta ^{2}_{4} \over \eta ^4} \right) ({\rm i}\ell) \,,
\cr}
$$
where
$$
{\rm Vol} = \sqrt{{\rm det} (g/\alpha')}
$$
denotes the volume of the four dimensional internal manifold.
Extracting the leading contributions to the tadpoles, 
one can show that $\tilde{\cal K}$
effectively involves states with coefficients that are perfect squares
as required by the consistency of the two-dimensional conformal field
theory in the presence of boundaries and/or crosscaps \refs{\bs,\fps}
\eqn\ktmz{
\eqalign{
\tilde{\cal K}^{(r)}_{0} =& {2^5 \over 4} \Biggl[ 
(Q_O \phi_O + Q_V \phi_V ) \left( \sqrt{{\rm Vol}} + {2^{-r/2} \over
\sqrt{{\rm Vol}}} \right)^2 +
\cr
&\qquad + 
(Q_O \phi_V + Q_V \phi_O ) \left( \sqrt{{\rm Vol}} - {2^{-r/2} \over
\sqrt{{\rm Vol}}} \right)^2 \Biggr] \,.
\cr}
}
Before turning to the open sector, let us pause for a moment and comment
on the action of the NS-NS antisymmetric tensor on the twisted
sector. Although it is evident that a non-vanishing $B$-flux modifies
the lattice sum, projecting on suitable winding states that satisfy \quant,
less obvious is the fact that it alters the structure of the twisted
sector, that does not depend on the moduli defining the
size and the shape of the lattice, and {\it a priori} does not
know anything about the $B_{ab}$ background. Still, simply imposing
the ``rule of perfect squares'' on the crosscap-to-crosscap reflection
coefficients for closed-string states one would have discovered that, 
effectively, $B_{ab}$ alters the $\Omega$-projection on the
fixed points.
In this way, one can easily obtain the correct
parametrisation, even without resorting to
a geometrical picture of the orbifold model, an option clearly of interest,
for instance, for 
asymmetric orbifolds. The same procedure
can then be applied to the open-string annulus and M{\"o}bius amplitudes
that, in the transverse channel, have to satisfy similar constraints.

We can now proceed to the construction of the open descendants,
introducing the open sector. From the torus amplitude \torusmz\ and
from our knowledge of the structure of the fixed points, we can
write the contributions of massless states to the annulus amplitude
\eqn\atmz{
\eqalign{
\tilde{\cal A}^{(r)}_{0} =& {2^{-5}\over 4} \Biggl\{
(Q_O \phi_O + Q_V \phi_V ) \left( 2^{r/2} \sqrt{{\rm Vol}} \, I
_{N} + {1 \over \sqrt{{\rm Vol}}} \sum_{i=1}^{2^{4-r}} I
^{i}_{D} \right)^2 +
\cr
& \quad \qquad 
+ (Q_O \phi_V + Q_V \phi_O ) \left( 2^{r/2} \sqrt{{\rm Vol}} \, I
_{N} - {1 \over \sqrt{{\rm Vol}}} \sum_{i=1}^{2^{4-r}} I
^{i}_{D}  \right)^2 +
\cr
& \quad \qquad +16 \sum_{i=1}^{2^{4-r}} \Biggl[ 
(Q_S \phi_S + Q_C \phi_C ) \left( 2^{(r-4)/2}\, R_N   - R
^{i}_{D} \right)^2 +
\cr
&\qquad \quad \qquad \qquad \qquad 
+ (Q_S \phi_C + Q_C \phi_S ) \left( 
2^{(r-4)/2}\, R_N  + R ^{i}_{D} \right)^2
\Biggr] \Biggr\} \,,
\cr}
}
where we have already related the
boundary-to-boundary reflection coefficients to the Chan-Paton
multiplicities. Here $I$ denotes the sum of Chan-Paton charges, while
$R$ parametrises the orbifold-induced gauge symmetry breaking. The
indices $N$ and $D$ refer to Neumann (D9-branes) and Dirichlet
(D5-branes) charges, respectively. Introducing the contributions from
momentum and (projected) winding massive states, one obtains the following
expressions for the annulus amplitude in the transverse channel
$$
\eqalign{
\tilde{\cal A}^{(r)} = {2^{-5}\over 4} \Biggl\{ & (Q_O + Q_V )({\rm
i}\ell)
\Biggl[ 2^{r-4} {\rm Vol} \,\, I ^{2}_{N} \sum_{\epsilon =0,1} \sum_n
{ ({\rm e}^{-2\pi\ell})^{{1\over 4\alpha '} n^{{\rm T}} g n} {\rm
e}^{{2 {\rm i}\pi \over \alpha '} n^{{\rm T}} B\epsilon} \over \eta ^4
({\rm i}\ell )} +
\cr
&\qquad\qquad\qquad + {1\over {\rm Vol}} \sum_{i,j=1}^{2^{4-r}} I_{D}^{i}
I_{D}^{j}
\sum_m {({\rm e}^{-2\pi\ell})^{{\alpha '\over 4} m^{{\rm T}} g^{-1} m}
{\rm e}^{2{\rm i}\pi m^{{\rm T}} (x^i - x^j)} \over \eta ^4 ({\rm i}
\ell ) } \Biggr] +
\cr
&+ 2\times 2^{r/2} (Q_O - Q_V) ({\rm i} \ell) \left(
{\vartheta_{3}^{2} \vartheta_{4}^{2} \over \eta ^4}\right) ({\rm
i}\ell)\sum_{i=1}^{2^{4-r}} I_N I_{D}^{i} +
\cr
&+ 4 (Q_S + Q_C) ({\rm i}\ell) \left({\vartheta_{2}^{2}
\vartheta_{3}^{2} \over \eta ^4}\right) ({\rm i}\ell) \left[ R_{N}^{2}
+
\sum_{i=1}^{2^{4-r}} (R_{D}^{i})^2 \right] +
\cr
&-2\times 2^{r/2} (Q_S -Q_C) ({\rm i}\ell) \left({\vartheta_{2}^{2}
\vartheta_{4}^{2} \over \eta ^4} \right) ({\rm i}\ell )
\sum_{i=1}^{2^{4-r}} R_N R_{D}^{i} \Biggr\}
\cr}
$$
and in the direct channel
\eqn\ad{
\eqalign{
{\cal A}^{(r)} = {\textstyle{1\over 4}} \Biggl\{ & (Q_O + Q_V)
(\sqrt{q}) \Biggl[ 2^{r-4} I_{N}^{2} \sum_{\epsilon =0,1} \sum_m
{q^{{\alpha '\over 2} (m + {1\over \alpha '} B\epsilon )^{{\rm T}}
g^{-1} (m + {1\over \alpha '} B\epsilon )} \over \eta ^4 (\sqrt{q})} +
\cr
&\qquad\qquad \qquad\quad
+ \sum_{i,j=1}^{2^{4-r}} I_{D}^{i} I_{D}^{j} \sum_n 
{q^{{1\over 2\alpha '} (n + {1\over 2} (x^i - x^j))^{{\rm T}} g (n +
{1\over 2} (x^i - x^j))} \over \eta ^4 (\sqrt{q})} \Biggr] +
\cr
&+ (Q_O - Q_V)(\sqrt{q}) \left( {\vartheta_{3}^{2} \vartheta_{4}^{2}
\over \eta ^4}\right)(\sqrt{q}) \left[ R^{2}_{N} +
\sum_{i=1}^{2^{4-r}} (R_{D}^{i})^2 \right] +
\cr
&+{2^{r/2}\over 2} (Q_S +Q_C)(\sqrt{q}) \left({\vartheta_{2}^{2}
\vartheta_{3}^{2} \over \eta ^4}\right) (\sqrt{q}) \sum_{i=1}^{2^{4-r}}
I_N I_{D}^{i} +
\cr
&+{2^{r/2}\over 2} (Q_S -Q_C)(\sqrt{q}) \left({\vartheta_{2}^{2}
\vartheta_{4}^{2} \over \eta ^4}\right) (\sqrt{q}) \sum_{i=1}^{2^{4-r}}
R_N R_{D}^{i} \Biggr\} \,.
\cr}
}
From this last expression, one can anticipate that states in the
twisted sector will have multiplicities related to the rank of
the NS-NS antisymmetric tensor. Once more, this non-trivial feature 
emerges naturally, imposing that the boundary-to-boundary reflection
coefficients be perfect squares.

To conclude the construction of the open descendants, one has to add
the contribution of the M{\"o}bius amplitude. From \ktmz\ and \atmz, we
can deduce the terms at the origin of the lattices
$$
\eqalign{
\tilde{\cal M}^{(r)}_{0} = - {\textstyle{2\over 4}}
\Biggl[ &
(\hat Q _O \hat \phi _O + \hat Q_V \hat\phi_V) \left( \sqrt{{\rm Vol}}
+ {2^{-r/2}\over \sqrt{{\rm Vol}}}\right) 
\left( 2^{r/2} \sqrt{{\rm Vol}} I_N + {1\over
\sqrt{{\rm Vol}}} \sum_{i=1}^{2^{4-r}} I_{D}^{i} \right) +
\cr
+& (\hat Q _O \hat \phi _V + \hat Q_V \hat\phi_O) \left( \sqrt{{\rm Vol}}
- {2^{-r/2}\over \sqrt{{\rm Vol}}}\right) 
\left( 2^{r/2} \sqrt{{\rm Vol}} I_N - {1\over
\sqrt{{\rm Vol}}} \sum_{i=1}^{2^{4-r}} I_{D}^{i} \right) \Biggr]
\cr}
$$
that, together with momentum and winding massive states, arrange in
$$
\eqalign{
\tilde{\cal M}^{(r)} = - {\textstyle{2\over 4}} \Biggl\{ & (\hat Q_O +
\hat Q_V ) ({\rm i}\ell + {\textstyle{1\over 2}}) \Biggl[
2^{(r-4)/2}{\rm Vol} \, I_N \sum_{\epsilon =0,1} \sum_n
{({\rm e}^{-2\pi\ell})^{{1\over \alpha '} n^{{\rm T}} g n} {\rm
e}^{{2{\rm i}\pi\over\alpha '} n^{{\rm T}} B\epsilon} \gamma_\epsilon
\over \hat\eta ^4 ({\rm i}\ell + {\textstyle{1\over 2}})} +
\cr
&\qquad\quad +{2^{-2}\over {\rm Vol}} \sum_{i=1}^{2^{4-r}} I_{D}^{i}
\sum_{\epsilon = 0,1}\sum_m {({\rm e}^{-2\pi\ell})^{\alpha ' (m +
{1\over \alpha '} B \epsilon)^{{\rm T}} g^{-1} (m + {1\over \alpha '}
B \epsilon)} \tilde\gamma_\epsilon \over \hat\eta ^4 ({\rm i}\ell +
{\textstyle{1\over 2}})} \Biggr] +
\cr
&+ (\hat Q_O - \hat Q_V ) ({\rm i}\ell +{\textstyle{1\over 2}} )\left(
{\hat\vartheta_{3}^{2} \hat\vartheta_{4}^{2} \over\hat\eta ^4}\right)
({\rm i}\ell +{\textstyle{1\over 2}}) \left( I_N + \sum_{i=1}^{2^{4-r}}
I_{D}^{i} \right) \Biggr\} \,.
\cr}
$$
Notice that both the momentum and the winding sums in the 
M{\"o}bius amplitude depend on $B_{ab}$. This is expected, since in the
transverse channel $\tilde{\cal K}$, $\tilde{\cal A}$ and $\tilde{\cal
M}$ must factorise. There is another very important point to
stress. The M{\"o}bius amplitude involves signs $\gamma_\epsilon$ and
$\tilde\gamma_\epsilon$ that give a correct normalisation to the
projectors onto states that satisfy \quant, 
similarly to what happens in toroidal compactifications. A $P$ modular
transformation then gives the M{\"o}bius amplitude in the direct channel
\eqn\md{
\eqalign{
{\cal M}^{(r)} =-{\textstyle{1\over 4}} \Biggl\{ & (\hat Q_O + \hat
Q_V ) (-\sqrt{q}) \Biggl[ 2^{(r-4)/2} I_N \sum_{\epsilon =0,1} \sum_m
{q^{{\alpha ' \over 2} (m + {1\over \alpha '} B\epsilon )^{{\rm T}}
g^{-1} (m + {1\over \alpha '} B\epsilon)} \gamma_\epsilon \over
\hat\eta ^4 (-\sqrt{q})} +
\cr
&\qquad\quad + 2^{-2} \sum_{i=1}^{2^{4-r}} I_{D}^{i} \sum_{\epsilon
=0,1} \sum_n {q^{{1\over 2\alpha '} n^{{\rm T}} g n} {\rm e}^{{2{\rm
i}\pi\over \alpha '} n^{{\rm T}} B\epsilon}\tilde\gamma_\epsilon \over
\hat\eta ^4 (-\sqrt{q})}\Biggr] +
\cr
& - (\hat Q_O - \hat Q_V ) (-\sqrt{q}) \left( {\hat\vartheta_{3}^{2}
\hat\vartheta_{4}^{2} \over \hat\eta ^4}\right) (-\sqrt{q}) \left( I_N
+
\sum_{i=1}^{2^{4-r}} I_{D}^{i} \right)\Biggr\}
\cr}
}
that completes the construction of the open descendants.

%%%%%%%%%%%%%%%%%%%%%%%%%%%%%%%%%%%%%%%%%%%%%%%%%%%%%%%%%%%%%%%%%%%%%%%%%%

\newsec{Tadpoles, discrete Wilson lines and anomalies}

We are now ready to extract the massless contributions to
the transverse-channel amplitudes, and thus to
read the tadpoles for the unphysical
states. The tadpole conditions are equivalent to those of
the standard $Z_2$ orbifold without $B_{ab}$ \refs{\ps,\gp}, aside from
the fact that now the total size of the gauge group is reduced
according to the rank of the NS-NS antisymmetric tensor. One thus gets
$$
\sqrt{{\rm Vol}} \, \left( 2^5 - 2^{r/2}\, 
I_N \right) 
\,
\pm \, {1\over \sqrt{{\rm Vol}}} \left( 2^{5-r/2} -
\sum_{i=1}^{n_+ - n_-}
I_{D}^{i} \right) = 0 
$$
from untwisted states, and
$$
2^{(r-4)/2}\, R_N - 
R_{D}^{i} = 0 \,, \qquad \quad {\rm for}\quad
i = 1 , \ldots , 2^{4-r} \,,
$$
from twisted states.

We have now all the ingredients needed to extract the spectrum of
massless excitations. We have already described the closed spectrum in
the previous section. It comprises the ${\cal N} =(1,0)$ supergravity
multiplet coupled to $1+n_-$ tensor multiplets and $4+n_+$
hypermultiplets. In order to extract the massless open spectrum, 
we expand the
amplitudes \ad\ and \md\ to lowest order in $q$, obtaining
$$
\eqalign{
{\cal A}^{(r)}_{0} \sim & {\textstyle{1\over 4}}
\left( I_{N}^{2} + R_{N}^{2}
+\sum_{i=1}^{2^{4-r}} (I_{D}^{i})^2 +
(R_{D}^{i} )^2
\right)\, (Q_O \phi_O + Q_V \phi_V ) +
\cr
&+ {\textstyle{1\over 4}}
\left( I_{N}^{2} - R_{N}^{2} +\sum_{i=1}^{2^{4-r}} 
(I_{D}^{i})^2 - ( R_{D}^{i} )^2 
\right)\, (Q_O \phi_V + Q_V \phi_O ) +
\cr
& + {2^{r/2} \over 2}\,\sum_{i=1}^{2^{4-r}} \left( I_N \, I_{D}^{i} +
R_N \, R_{D}^{i} \right) (Q_S \phi_S + Q_C \phi_C) +
\cr
& + {2^{r/2}\over 2}\,\sum_{i=1}^{2^{4-r}} \left( I_N \,
I_{D}^{i} - R_N \, R_{D}^{i} \right) (Q_S \phi_C + Q_C \phi_S) 
\cr}
$$
for the annulus amplitude, and
$$
\eqalign{
{\cal M}^{(r)}_{0} \sim &  - {\textstyle{1\over 4}}
(\hat Q_O + \hat Q_V) (\hat\phi_O +\hat\phi_V) \left( \gamma_0 \, I_N
+ \tilde\gamma_0 \sum_{i=1}^{2^{4-r}} I_{D}^{i} \right) +
\cr
& \quad + {\textstyle{1\over 4}}
(\hat Q_O - \hat Q_V) (\hat\phi_O -\hat\phi_V) \left(
I_N +
\sum_{i=1}^{2^{4-r}} I_{D}^{i} \right)
\cr}
$$
for the M{\"o}bius amplitude. In order to read the spectrum from the
above expressions, we must still introduce an explicit
parametrisation of $I_\alpha$ and $R_\alpha$ in terms of Chan-Paton
multiplicities. It is naturally obtained demanding a consistent particle
interpretation of the physical spectrum, {\it i.e.} imposing that the
M{\"o}bius amplitude symmetrises correctly the annulus amplitude. It is
then evident that the signs $\gamma_0$ and
$\tilde\gamma_0$ present in ${\cal M}$ play a crucial role. In
order to respect the reality of the annulus amplitude and the
structure of the transverse channel, they have to coincide, 
{\it i.e.} $\gamma_0
= \tilde\gamma_0$, and then we are left with two
possibilities: $\gamma_0 = \tilde\gamma_0 = +1$ and $\gamma_0 =
\tilde\gamma_0 = -1$. In the former case, the M{\"o}bius amplitude
contains at the massless level only the untwisted hypermultiplet. As a
result, the gauge group is unitary and one is led to the following
parametrisation in terms of complex Chan-Paton charges:
$$
\eqalign{
I_N &= N+\bar N \,,
\cr
I_{D}^{i} &= D^i+\bar D^i \,,
\cr}
\qquad
\eqalign{
R_N &= {\rm i}\, (N-\bar N) \,,
\cr
R_{D}^{i} &= {\rm i}\, (D^i-\bar D^i) \,.
\cr}
$$
This is consistent with the well known result for the $T^4 /Z_2$ orbifold
\refs{\ps,\gp} with vanishing $B$-field, for which the signs
$\gamma_0$ and $\tilde\gamma_0$ are uniquely fixed by the tadpole
conditions. The massless spectrum then comprises non-abelian
vector multiplets with gauge group 
$$
G_{{\rm CP}} = \left. {\rm U} (2^{4-r/2})\right|_{99} 
\otimes \left.{\rm U} (2^{4-r/2}) \right|_{55}
$$
and additional charged hypermultiplets in the representations 
$$
2\, (A ; 1) \oplus 2\, (1; A) \oplus 2^{r/2} \, (F;F) \,,
$$
where $F$ ($A$) denotes the fundamental (antisymmetric) representation. 
Here and in the following
we have decided to confine all the D5-branes to the same
fixed point. In general, however, it is possible to distribute them at 
different
fixed points or at any generic positions in the internal manifold, thus
breaking the gauge group to a product of unitary and/or symplectic subgroups.  
In table 1 we summarise the massless spectra for the
various choices of $r=0,2,4$.

\vbox{
\vskip 18pt
\settabs 8\columns
\hrule
\vskip 4pt
\+ 
\ $r$ & $n_{T}^{c}$ & $n_{H}^{c}$ & $G_{{\rm CP}}$ & & open hypermultiplets
\cr
\vskip 4pt
\hrule
\vskip 4pt
\+
\ 0 &1 & 20 & ${\rm U} (16)_{99} \otimes {\rm U} (16)_{55}$ & & $2\,
(120;1)\oplus 2\,(1;120)\oplus (16;16)$
\cr
\+
\ 2 &5 & 16 & ${\rm U} (8)_{99} \otimes {\rm U} (8)_{55}$ & & $2\,
(28;1)\oplus 2\,(1;28)\oplus 2\, (8;8)$
\cr
\+
\ 4 &7 & 14 & ${\rm U} (4)_{99} \otimes {\rm U} (4)_{55}$ & & $2\,
(6;1)\oplus 2\,(1;6)\oplus 4\, (4;4)$
\cr
\cleartabs
\vskip 4pt
\hrule
\vskip 12pt
{\ninepoint \centerline{{\bf Table 1.} Massless spectrum for $\gamma_0 =
\tilde\gamma _0 =+1$.}}
\vskip 18pt
}

When $r\not= 0$, one has the additional choice $\gamma_0 =
\tilde\gamma_0 = -1$. In this case, the M{\"o}bius amplitude contains the
massless vector with a positive sign, and therefore the contribution
of the annulus amplitude is 
symmetrised\footnote{${}^{\S}$}{See also \kakusp\ for 
a previous discussion about the appearance of symplectic gauge groups in
the $T^4/Z_2$ orbifold.}. 
The gauge groups are thus
symplectic, and call for the following
parametrisation in terms of real Chan-Paton charges:
$$
\eqalign{
I_N &= N_1+N_2 \,,
\cr
I_{D}^{i} &= D_{1}^{i}+ D_{2}^{i} \,,
\cr}
\qquad
\eqalign{
R_N &= N_1- N_2 \,,
\cr
R_{D}^{i} &= D_{1}^{i}- D_{2}^{i} \,.
\cr}
$$
As a result, one finds a generic gauge group of the form
$$
G_{{\rm CP}} = \left. {\rm Sp} (2^{4-r/2}) \otimes {\rm Sp} (2^{4-r/2})
\right|_{99} \otimes \left. {\rm Sp} (2^{4-r/2}) \otimes {\rm Sp} 
(2^{4-r/2}) \right|_{55}
$$
with charged hypermultiplets in bi-fundamentals. In table 2 we
summarise the resulting massless spectra for the choices $r=2,4$.

\vbox{
\vskip 18pt
\settabs 8\columns
\hrule
\vskip 4pt
\+ 
\ $r$ & $n_{T}^{c}$ & $n_{H}^{c}$ & $G_{{\rm CP}}$ & & open hypermultiplets
\cr
\vskip 4pt
\hrule
\vskip 4pt
\+
\ 2 &5 & 16 & ${\rm Sp}(8)^{2}_{99} \otimes {\rm Sp}(8)^{2}_{55}$ & & 
$(8,8;1,1)\oplus (1,1;8,8)$
\cr
\+ 
& & & & & $\quad\oplus (8,1;8,1)\oplus (1,8;1,8)$
\cr
\+
\ 4 &7 & 14 & ${\rm Sp} (4)^{2}_{99} \otimes {\rm Sp} (4)^{2}_{55}$ & & 
$(4,4;1,1)\oplus (1,1;4,4)$
\cr
\+
& & & & & $\quad\oplus 2\, (4,1;4,1) \oplus 2\, (1,4;1,4)$
\cr
\cleartabs
\vskip 4pt
\hrule
\vskip 12pt
{\ninepoint \centerline{{\bf Table 2.} Massless spectrum for $\gamma_0 =
\tilde\gamma _0 =-1$.}}
\vskip 18pt
}

It is not surprising to realize that all these models had been already
discovered in \refs{\bs ,\gepner}, using the rational algorithm. 
In fact, group lattices naturally involve a non-vanishing
quantised $B$-field given explicitly by the adjacency matrix of the
associated algebra. In particular, the $Z_2$ models discussed
in \refs{\bs ,\gepner} were built starting from a toroidal
compactification on the ${\rm SO} (4)^2$ lattice and on the ${\rm SO}
(8)$ lattice. The latter has a $B_{ab}$ with $r=2$ and, indeed, the
spectrum of the open descendants associated to the (geometrical)
diagonal modular invariant partition function is exactly the one
reported in table 2. It is quite surprising, however, to see how the
second model is obtained in the rational construction. Actually, it is
related to a different modular invariant combination of characters
for the same ${\rm SO} (8)$ lattice that, as we have
seen, involves naively a rank-two antisymmetric tensor. Actually, this
is true for the geometrical combination of characters corresponding to
the diagonal (or, better, charge-conjugation) modular invariant
$$
|O_8|^2 + |V_8|^2 + |S_8|^2 + |C_8|^2 \, ,
$$
while all other non-geometrical choices, as for instance
$$
|O_8|^2 + |V_8|^2 + S_8 \bar C_8 + C_8 \bar S_8 \,,
$$
effectively increase the rank of the $B_{ab}$ matrix.

Actually, it was shown in \bs\ that the models in table 1
can also be obtained in the rational construction. They involve
the introduction of discrete Wilson lines in the M{\"o}bius amplitude,
that have the effect to modify the $P$ transformation. In order to
illustrate this phenomenon,
let us consider the models with $r=2$. In the rational model, the
transverse-channel M{\"o}bius amplitude is \bs
$$
\tilde{\cal M} = - \alpha (\hat Q _O \hat O_4 \hat O_4 - \hat Q_V
\hat V_4 \hat V_4 ) \,.
$$
Since $P$ acts as $\sigma_1$ on $(\hat O _4 , \hat V_4)$, with
$\sigma_1$ the first Pauli matrix, it is evident
that the direct-channel amplitude
$$
{\cal M} = \alpha (\hat Q _O \hat O_4 \hat O_4 - \hat Q_V
\hat V_4 \hat V_4 )
$$
calls for symplectic gauge groups. Actually, this is not the
only possible choice. Compatibly with the supercurrent, one may
introduce discrete Wislon lines, changing the definition of the
internal characters
$$
\eqalign{
\hat O_4 &= \hat O _2 \hat O_2 - \hat V_2 \hat V_2 \ \to\ \hat O
^{\prime}_{4} = \hat O_2 \hat O_2 + \hat V_2 \hat V_2 \,,
\cr
\hat V_4 &= \hat O _2 \hat V_2 + \hat V_2 \hat O_2 \ \to \ \hat V
^{\prime}_{4} = \hat O_2 \hat V_2 - \hat V_2 \hat O_2 \,,
\cr}
$$
and thus affecting the $P$ transformation, whose action on the
primed characters $(\hat O^{\prime}_{4} \,,\, \hat V ^{\prime}_{4} )$
is now represented by the Pauli matrix $\sigma_3$. As a result, the new
$$
\tilde{\cal M} ' = - \alpha  (\hat Q _O \hat O^{\prime}_{4} \hat
O^{\prime}_{4}  - \hat Q_V
\hat V^{\prime}_{4} \hat V^{\prime}_{4} )
$$
maps to
$$
{\cal M} ' = - \alpha (\hat Q _V \hat O^{\prime}_{4} \hat
O^{\prime}_{4}  - \hat Q_O
\hat V^{\prime}_{4} \hat V^{\prime}_{4} )
$$
and calls for unitary gauge groups. We are thus led 
to identify the action of discrete Wilson lines for non-zero 
$B_{ab}$ with the signs
$\gamma_\epsilon$ and $\tilde\gamma_\epsilon$ present in the
irrational construction.

As a side remark, let us recall how the ${\rm U} (16) \otimes {\rm U}
(16)$ model was originally built in \bs . It originates from
the compactification on the ${\rm SO} (4)^2$ lattice, once the residual
internal global ${\rm SO} (2)^4$ symmetry is broken, resorting to eight
copies of the Ising model. As a result, all characters become real and
contribute to the transverse annulus amplitude, thus enhancing the
overall size of the Chan-Paton gauge group. At this point, as in the
previous case, the introduction of discrete Wilson lines turns each
pair of Sp gauge groups into a single unitary group, thus yielding
the ${\rm U} (16) \otimes {\rm U} (16)$ model. 

To close this section, let us comment on the cancellation of the residual
anomalies for these models. As usual, tadpole cancellations guarantee
the vanishing of the irreducible ${\rm tr} R^4$ and 
${\rm tr} F^{4}_{\alpha}$ terms. However, this is not in general
sufficient, and some other mechanisms are required in order to
compensate the reducible terms. In $D=10$ this is the familiar 
Green-Schwarz mechanism, that involves the universal 2-form present
in the spectrum \gs. This mechanism still applies to any
compactification with a single 2-form, for instance for the K3
reduction of the heterotic
string, or for the $T^4/Z_2$ reduction of the type-I string
with a vanishing background for
the NS-NS $B_{ab}$. However, when additional tensor fields are present (a
generic feature of six-dimensional open-string vacua), the
cancellation of gauge and gravitational anomalies 
requires a generalised Green-Schwarz mechanism \ggsm. In
fact, for $n_T$ tensor multiplets the residual anomaly polynomial
takes the form\footnote{${}^\ddagger$}{In this expression, for $r=s=0$ the
indices $x$ and $y$ refer to the contribution of the gravitational
term ${\rm tr} R^2$. In these supersymmetric models, the self-dual 2-form 
present in the gravitational multiplet remove
the residual gravitational and mixed anomalies.}
$$
{\cal I}_8 = - \sum_{x,y} c^{r}_{x} c^{s}_{y} \eta_{rs} {\rm tr}_x F^2
{\rm tr}_y F^2 \,,
$$
with the $c$'s a collection of constants and $\eta$ the Minkowski
metric for ${\rm SO} (1,n_T)$. Accordingly, the field strengths of the
2-forms include suitable combinations of (Yang-Mills and
gravitational) Chern-Simons forms, and the kinetic term for the gauge
vectors involves couplings to the scalars $v_r$ in the tensor
multiplets:
$$
e^{-1}\,{\cal L} \sim - {\textstyle{1\over 2}} v_r c^{rz} {\rm tr}_z
F_{\mu \nu} F^{\mu \nu}\,,
$$
with corresponding generic strong-coupling singularities \ggsm\ that
may be related to the appearance of tensionless strings \tensionless.
Actually, it is not true that the presence of additional tensor
fields in the spectrum automatically calls for a generalised
Green-Schwarz mechanism involving all of them. For instance, let us consider
the models reported in table 1. In this case the residual anomaly
polynomial is 
\eqn\anom{
\eqalign{
{\cal I}_{8}^{(r)} =& - {2^{2/r} \over 16} \left( 2^{-r/2} {\rm tr}
R^2 - {\rm tr} F_{1}^{2} - {\rm tr} F_{2}^{2} \right)^2 +
{2^{r/2} \over 16} \left( {\rm tr} F_{1}^{2} - {\rm tr} F_{2}^{2}
\right)^2 +
\cr
&+ {2^{r/2} \over 6}\, {\rm tr} F_1 \, \left( {\textstyle{1\over 16}}
{\rm tr} R^2 \, {\rm tr} F_2 - {\rm tr} F_{2}^{3} \right) +
{2^{r/2} \over 6} \, {\rm tr} F_2 \, \left( {\textstyle{1\over 16}}
{\rm tr} R^2 \, {\rm tr} F_1 - {\rm tr} F_{1}^{3} \right) \,.
\cr}
}
From this expression, one can read that, besides the self-dual 2-form 
present in the
supergravity multiplet, only the untwisted antiself-dual 2-form is
involved in the Green-Schwarz mechanism (aside from twisted
scalar fields responsible for the cancellation of abelian
anomalies). This has been explicitly checked in
\refs{\treshold,\serone,\bst} for the $r=0$ case, where this result is
expected since only the untwisted tensor multiplet is
available. Nonetheless, one can use these results to justify the
structure of \anom\ for generic $r$. 
In fact, for the case $\gamma_0 = \tilde\gamma_0 =+1$, the explicit
value of the constants $c$ is\footnote{${}^\star$}{Here we are
only interested in the dependence of the constants on the Chan-Paton
charges. For a proper normalisation, we refer to \treshold.} \treshold
$$
c \sim (N-\bar N) \, ,
$$
as can be seen turning on a background magnetic field. 
Thus, due to the numerical identification of conjugate charges, there
is effectively no coupling between twisted scalars and vectors and, as
a result, only a pair of tensors is involved in the
generalised Green-Schwarz mechanism. The situation changes drastically
for the choice $\gamma_0 =\tilde\gamma_0 =-1$. In this case, each ${\rm
U} (N)$ group splits in two different symplectic groups ${\rm Sp} (N_1)
\otimes {\rm Sp} (N_2)$, and the constants $c$ associated to the
$\alpha$-th factor in the Chan-Paton gauge group are now given by
$$
c\sim N_\alpha \,.
$$
Thus, they do not vanish, as expected from the corresponding
expressions for the anomaly polynomials:
$$
\eqalign{
{\cal I}_8^{\prime\, (r)} =& - {2^{r/2}\over 64} \left( 2^{(2-r)/2}\,
{\rm tr} R^2 -
{\rm tr} F_{1}^{2} - {\rm tr} F_{2}^{2} - {\rm tr} F_{3}^{2} -
{\rm tr} F_{4}^{2} \right)^2 +
\cr
&+ {2^{r/2}\over 64}
 \left( 
{\rm tr} F_{1}^{2} + {\rm tr} F_{2}^{2} - {\rm tr} F_{3}^{2} -
{\rm tr} F_{4}^{2} \right)^2 +
\cr
& + {2^{r/2}\over 64} (2 - 2^{(r-2)/2})
\left( 
{\rm tr} F_{1}^{2} - {\rm tr} F_{2}^{2} + {\rm tr} F_{3}^{2} -
{\rm tr} F_{4}^{2} \right)^2 +
\cr
&+ 
{2^{r/2}\over 64} (4 - 2^{(r-2)/2})
\left( 
{\rm tr} F_{1}^{2} - {\rm tr} F_{2}^{2} - {\rm tr} F_{3}^{2} +
{\rm tr} F_{4}^{2} \right)^2 
\cr}
$$
that clearly involve twisted antiself-dual 2-forms.

%%%%%%%%%%%%%%%%%%%%%%%%%%%%%%%%%%%%%%%%%%%%%%%%%%%%%%%%%%%%%%%%%%%%%%%%

\newsec{The $T^6 /Z_3$ orbifold with a non-vanishing $B_{ab}$}

In this section we want to apply our previous results to the 
four-dimensional $Z$
orbifold, thus generalising \chiral\ (see also \kakuz ). 
Since the orbifold group does
not contain any generator that squares to the identity, it is natural
to expect that this construction does not present any new interesting 
features. As we
shall see, however, this is not the case, since some subtleties 
are indeed present
in the construction of the M{\"o}bius amplitude. In what follows 
we will use the notation introduced in \chiral . 

The building blocks for the $Z$ orbifold are the amplitudes ($\omega =
{\rm e}^{2 {\rm i} \pi /3}$, $\rho,\lambda =0,\pm 1$)
$$
\Xi_{\rho,\lambda} (q) = \left( {A_0 \chi_\rho + \omega ^\lambda A_+
\chi_{\rho -1} +
\bar\omega ^\lambda A_- \chi_{\rho +1} \over H^{3}_{\rho ,\lambda}}
\right) 
(q) \,,
$$
where
$$
\eqalign{
H_{0 ,\lambda} (q) &= q^{{1\over 12}} \prod_{n=1}^{\infty} (1-\omega
^\lambda q^n) (1- \bar\omega ^\lambda q^n ) \,,
\cr
H_{+ ,\lambda} (q) = H_{-,-\lambda} (q) &= 
{\textstyle{1\over \sqrt{3}}} \, q^{-{1\over 36}}
\prod_{n=0}^{\infty} (1 - \omega ^\lambda q^{n+{1\over 3}} )
(1-\bar\omega ^\lambda q^{n+{2\over 3}} )
\cr}
$$
denote the contribution of the internal bosons, while the level one
SU(3) characters $\chi_{0,\pm }$ and the
combinations\footnote{${}^{*}$}{Here $(O_2, V_2 , S_2 , C_2)$ are
level one SO(2) characters while $\xi_m$ ($m=-5,\ldots ,6$) are the 12
characters of the ${\cal N}=2$ superconformal model with $c=1$,
equivalent to the rational torus at radius $R=\sqrt{12}$.}
$$
A_\rho = V_2 \, \xi_{4\rho} + O_2 \, \xi_{4\rho+6} - S_2 \, \xi_{4\rho -3} -
C_2 \, \xi_{4\rho +3} 
$$
denote the contribution of the world-sheet fermions.
Thus, the torus amplitude for the type IIB string is
$$
\eqalign{
{\cal T} =& {\textstyle{1\over 3}} \Biggl[ \Xi_{0,0} \bar \Xi_{0,0} 
\sum_{m,n} q^{{\alpha ' \over 4} p^{{\rm T}} g^{-1} p} \bar q
^{{\alpha ' \over 4} \tilde p ^{{\rm T}} g^{-1} \tilde p} 
+ \sum_{\lambda =\pm 1} \Xi_{0,\lambda} \bar \Xi _{0,-\lambda} +
\cr
& \quad + \sum_{\rho =\pm 1} \sum_{\lambda =0,\pm 1} \Xi_{\rho ,
\lambda} \bar \Xi_{-\rho ,-\lambda} \Biggr] \,.
\cr}
$$
The left and right momenta are given explicitly by \momenta , with 
the metric $g_{ab}$ pertaining to three orthogonal copies of a 
two-dimensional hexagonal lattice, as required by the $Z_3$ symmetry.

As in toroidal models, the Klein bottle amplitude is not modified by
the presence of the NS-NS antisymmetric tensor. It is given by \chiral
$$
{\cal K} = {\textstyle{1\over 6}} \, \Biggl[\Xi_{0,0} (q^2)  \sum_m
q^{{\alpha ' \over 2} m^{{\rm T}} g^{-1} m}  +
\sum_{\lambda =\pm 1} \Xi_{0,\lambda} (q^2) \Biggr]
$$
and, together with the torus amplitude, leaves at the massless level
the ${\cal N} =1$ supergravity multiplet, a universal chiral multiplet
and 9 additional chiral multiplets from the untwisted sector, as well
as 27 chiral multiplets from the twisted sectors.

The open-string sector now involves in the transverse-channel 
the projector on the winding states satisfying \quant , and
results in the following one-loop amplitudes:
$$
{\cal A} = {\textstyle{1\over 6}} \Biggl[ N^{2}_{0} \, 2^{r-6}\, \Xi_{0,0}
(\sqrt{q}) \sum_{\epsilon = 0,1} \sum_m q^{{\alpha ' \over 2}
(m + {1\over \alpha '} B\epsilon)^{{\rm T}} g^{-1} (m + {1\over
\alpha '} B\epsilon)}  + \sum_{\lambda = \pm 1}
N^{2}_{\lambda} \, \Xi_{0,\lambda} (\sqrt{q}) \Biggr]
$$
and
$$
\eqalign{
{\cal M} = - {\textstyle{1\over 6}} \Biggl[ & \delta_0 \, N_0 \,
2^{(r-6)/2} \, \hat\Xi _{0,0} (-\sqrt{q}) \sum_{\epsilon =0,1}
\sum_m q^{{\alpha ' \over 2} (m+{1\over \alpha '} B \epsilon )^{{\rm
T}} g^{-1} (m+ {1\over \alpha '} B\epsilon)} \gamma_\epsilon 
+
\cr
& + \sum_{\lambda =\pm 1} \delta_\lambda \, N_\lambda \, \hat \Xi
_{0,\lambda} \Biggr] \,,
\cr}
$$
where
$$
N_\lambda = n + \omega^\lambda m + \bar\omega^\lambda \bar m 
$$
are suitable combinations of the Chan-Paton multiplicities $n,\, m ,\,
\bar m$. The M{\"o}bius amplitude involves different kinds of signs
$\delta_\lambda$ and $\gamma_\epsilon$. The latter, by now familiar,
enforce a proper normalisation of the M{\"o}bius amplitude, whereas
the former are fixed by the tadpole conditions
$$
\eqalign{
n+m+\bar m &= 2^{5-r/2}\,\delta_0
\cr
n+\omega^\lambda m + \bar\omega^\lambda \bar m &= -4\,\delta_\lambda
\qquad\quad (\lambda =\pm 1)
\cr}
$$
whose solution requires $\delta_0 = +1$ and $\delta_\pm = +1$
($\delta_\pm = -1$) for $r=0,4$ ($r=2,6$). In order to better
appreciate the role of the signs $\delta_\pm $ and
$\gamma_\epsilon$, let us expand the open-string amplitudes keeping
only the massless contributions. One has
$$
{\cal A}_0 \sim \left( {n^2 \over 2} + m \bar m \right) \, A_0 \chi_0
\, + \left( {\bar m ^2 \over 2} + nm \right) \, A_+ \chi_- \, + \left(
{m^2 \over 2} + n\bar m \right) \, A_- \chi_+ 
$$
for the annulus amplitude, and
$$
\eqalign{
{\cal M}_0 \sim -{\textstyle{1\over 6}} \Biggl\{ & \Bigl[ n (\gamma_0
+ \delta_+ + \delta_-) + m (\gamma_0 + \omega \delta_+ +\bar\omega
\delta_- ) + \bar m (\gamma_0 +\bar\omega \delta_+ + \omega \delta_-)
\Bigr] \hat A _0 \hat \chi_0 +
\cr
&+ \Bigl[ n (\gamma_0 
+ \bar\omega \delta_+ + \omega \delta_-) + m (\gamma_0 + \delta_+ +
\delta_- ) + \bar m (\gamma_0 +\omega \delta_+ + \bar\omega \delta_-)
\Bigr] \hat A _+ \hat \chi_- +
\cr
&+ \Bigl[ n (\gamma_0
+\omega \delta_+ + \bar\omega \delta_-) + m (\gamma_0 + \bar\omega
\delta_+  +\omega\delta_- ) + \bar m (\gamma_0 +\delta_+ + \delta_-)
\Bigr] \hat A _- \hat \chi_+ \Biggr\}
\cr}
$$
for the M{\"o}bius amplitude. It is then evident that, in order to have 
a consistent particle interpretation, the signs $\gamma_0$ and
$\delta_\pm $ are to be the same. Therefore, for the $Z$ orbifold one
does not have the option to introduce discrete Wilson lines for a given
$B_{ab}$. They are uniquely fixed by the rank of the NS-NS
antisymmetric tensor. Some of the gauge groups of reduced
rank in \chiral\ should thus 
be corrected, according to the following table:
\vbox{
\vskip 18pt
\settabs 8\columns
\hrule
\vskip 4pt
\+ 
& $r$ & & $G_{{\rm CP}}$ & & open chiral multiplets
\cr
\vskip 4pt
\hrule
\vskip 4pt
\+
& 0 & & ${\rm U} (12) \otimes {\rm SO} (8) $ & & $3\times (\overline{12},8)
\oplus 3\times (66,1)$
\cr
\+
& 2 & & ${\rm U} (4) \otimes {\rm Sp} (8)$ & & $3\times 
(\overline{4},8)\oplus 3\times (10,1)$
\cr
\+
& 4 & & ${\rm U} (4) $ & & $3\times 6$
\cr
\+
& 6 & & ${\rm Sp} (4) $ & & ---
\cr
\cleartabs
\vskip 4pt
\hrule
\vskip 12pt
{\ninepoint \centerline{{\bf Table 3.} Massless spectrum for the $T^6
/Z_3$ orbifold.}}
\vskip 18pt
}

%%%%%%%%%%%%%%%%%%%%%%%%%%%%%%%%%%%%%%%%%%%%%%%%%%%%%%%%%%%%%%%%%%%%%%%%

\newsec{Comments on brane supersymmetry breaking}

To conclude, let us briefly comment on the results of \ads. The
structure of the amplitudes presented in the previous sections is
uniquely fixed by the constraints of the  underlying two-dimensional
conformal field theory and by space-time supersymmetry. This
last requirement, for instance, prevents the introduction of signs in
the Klein bottle amplitude that are actually allowed by the 
crosscap constraint \fps. 
Actually, only one (non-trivial) choice is compatible with
supersymmetry
$$
{\cal K} = {\textstyle{1\over 4}} (Q_O + Q_V ) \left[ \sum_m {(-)^m
q^{{\alpha ' \over 2} m^{{\rm T}} g^{-1} m} \over \eta^4 } +
\sum_n {(-)^n q^{{1\over 2\alpha '} n^{{\rm T}} g n} \over \eta^4 }
\right]
+ {\textstyle{1\over 8}} (8-8) (Q_S + Q_C) {\vartheta^{2}_{2}
\vartheta^{2}_{3} \over \eta^4}
$$
that results in a model with 9 tensor multiplets, 12
hypermultiplets and {\it no} open strings. 

The nice observation of \ads\ consists in relaxing the
requirement of space-time supersymmetry, thus allowing for other
consistent
solutions of the crosscap constraint. In particular, they considered
the case in which the whole twisted sector is antisymmetrised in the
NS-NS sector and symmetrised in the R-R one, yelding an ${\cal N} =
(1,0)$ supersymmetric closed (bulk) spectrum with 17 tensor multiplets
and 4 hypermultiplets. The corresponding open spectrum is non-supersymmetric,
with non-abelian vectors in the adjoint of ${\rm SO} (16) \otimes {\rm
SO} (16) \otimes {\rm Sp} (16) \otimes {\rm Sp} (16)$ and additional
fermions and scalars in (anti)symmetric and bi-fundamental
representations. As a result of supersymmetry breaking, the M{\"o}bius
amplitude generates a one-loop cosmological constant supported on the
D$\bar 5$-branes and, consequently, a potential for 
the NS-NS moduli appears. 
This is precisely the opposite to what found in previous type I settings,
where supersymmetry is spontaneously broken in the bulk but can still be
present on the massless excitations \adds, or even on the whole massive 
spectrum \nonsusy, of suitable branes.
In the following we shall study the modifications induced
in this model by a non-vanishing antisymmetric tensor $B_{ab}$ present
in the compactification lattice. For the sake of brevity, we shall 
only present the
various amplitudes, referring to \ads\ and to the previous
sections for a detailed discussion of the subtleties of the
construction.

Aside from the torus amplitude \torus , one has the following
contributions:
$$
\eqalign{
{\cal K}^{(r)} =& {\textstyle{1\over 4}} (Q_O + Q_V)(q^2) \left[
\sum_m
{q^{{\alpha '\over 2} m^{{\rm T}} g^{-1} m} \over \eta^4 (q^2)} +
2^{-4} \sum_{\epsilon =0,1} \sum_n {q^{{1\over 2 \alpha '} n^{{\rm T}}
g n } {\rm e}^{{2 {\rm i}\pi \over \alpha '} n^{{\rm T}} B \epsilon}
\over \eta^4 (q^2 )} \right] +
\cr
& - {2^{(4-r)/2}\over 2} (Q_S +Q_C) (q^2) \left( {\vartheta^{2}_{2}
\vartheta^{2}_{3} \over \eta^4 }\right) (q^2 )
\cr}
$$
from the Klein bottle amplitude,
$$
\eqalign{
{\cal A}^{(r)} &= {\textstyle{1\over 4}} \Biggl\{ (Q_O + Q_V )
(\sqrt{q}) \Biggl[ 2^{r-4} \, I^{2}_{N} \, \sum_{\epsilon =0,1}
\sum_m {q^{{\alpha ' \over 2} (m + {1\over \alpha '} B \epsilon
)^{{\rm T}} g^{-1} (m + {1\over\alpha '} B \epsilon )} \over \eta^4
(\sqrt{q})} +
\cr
&\qquad \qquad \qquad + \sum_{i,j}^{2^{4-r}} I_{D}^{i} I_{D}^{j}
\sum_n {q^{{1\over 2\alpha '} (n + {1\over 2} (x^i - x^j ))^{{\rm T}}
g (n + {1\over 2} (x^i -x^j))} \over \eta^4 (\sqrt{q})} \Biggr] +
\cr
& \quad+ (Q_O - Q_V ) (\sqrt{q}) \left( {\vartheta^{2}_{3}
\vartheta^{2}_{4} \over \eta^4} \right) (\sqrt{q}) \left( R^{2}_{N} +
\sum_{i=1}^{2^{4-r}} (R^{i}_{D} )^2 \right) +
\cr
& \quad+ {2^{r/2} \over 2} (O_4 S_4 + V_4 C_4 - C_4 O_4 - S_4 V_4)
(\sqrt{q} ) \left( {\vartheta^{2}_{2} \vartheta^{2}_{3} \over \eta^4}
\right) (\sqrt{q}) \sum_{i=1}^{2^{4-r}} I_N I_{D}^{i} +
\cr
& \quad - {2^{r/2} \over 2} (O_4 S_4 - V_4 C_4 + C_4 O_4 - S_4 V_4 )
(\sqrt{q}) \left( {\vartheta^{2}_{2} \vartheta^{2}_{4} \over \eta^4}
\right) (\sqrt{q}) \sum_{i=1}^{2^{4-r}} R_N R^{i}_{D} \Biggr\}
\cr}
$$
from the annulus amplitude, and
$$
\eqalign{
{\cal M}^{(r)} &= -{\textstyle{1\over 4}} \Biggl\{ 2^{(r-4)/2} \, I_N\,
(\hat V_4 \hat O _4
+ \hat O_4 \hat V_4 - \hat C_4 \hat C_4 - \hat S_4 \hat S_4 )
(-\sqrt{q}) \times 
\cr
&\qquad\qquad\quad\times 
\sum_{\epsilon =0,1} \sum_m {q^{{\alpha ' \over 2} (m +
{1\over \alpha ' } B \epsilon )^{{\rm T}} g^{-1} (m + {1\over \alpha
'} B \epsilon )} \gamma_\epsilon \over \hat\eta ^4 (-\sqrt{q}) } +
\cr
&\quad - 2^{-2} \sum_{i=1}^{2^{4-r}} I_{D}^{i}\,
(\hat V_4 \hat O _4 + \hat O_4 \hat V_4 + \hat C_4 \hat C_4 + 
\hat S_4 \hat S_4 ) (-\sqrt{q})  \times 
\cr
&\qquad\qquad\quad\times \sum_{\epsilon =0,1} \sum_n 
{q^{{1\over 2 \alpha '} n^{{\rm T}} g n} {\rm e}^{{2{\rm i}\pi\over\alpha '}
n^{{\rm T}} B \epsilon} \tilde\gamma_\epsilon \over \hat\eta^4
(-\sqrt{q})} +
\cr
&\quad + I_N \, (\hat V_4 \hat O _4 - \hat O_4 \hat V_4 - \hat C_4
 \hat C_4 + \hat S_4 \hat S_4 ) (-\sqrt{q}) \left({\hat\vartheta^{2}_{3} 
\hat\vartheta^{2}_{4} \over \hat \eta ^4} \right) (-\sqrt{q}) +
\cr
&\quad - \sum_{i=1}^{2^{4-r}} I_{D}^{i} \, (\hat V_4 \hat O _4
- \hat O_4 \hat V_4 + \hat C_4 \hat C_4 - \hat S_4 \hat S_4)   
 (-\sqrt{q}) \left({\hat\vartheta^{2}_{3} \hat\vartheta^{2}_{4} \over
\hat\eta ^4}\right) (-\sqrt{q}) \Biggr\}
\cr}
$$
from the M{\"o}bius amplitude. As expected from our general
considerations, ${\cal M}$ involves the signs $\gamma_\epsilon$ and
$\tilde\gamma_\epsilon$, that play the same role as in the
supersymmetric case.

As in \ads , the cancellation of R-R tadpoles fixes both the overall size
of the Chan-Paton gauge group, that in this case, however, is reduced by a
factor $2^{r/2}$ both on the D9 and D$\bar 5$-branes, and the group breaking 
pattern. However, the untwisted NS-NS scalars develop non-vanishing
one-point functions, and thus a scalar potential arises.

Expanding the annulus and M{\"o}bius amplitudes to leading order in $q$,
one can easily appreciate the role of the signs $\gamma_\epsilon$ and
$\tilde\gamma_\epsilon$ that even in this case must coincide. One has
$$
\eqalign{
{\cal A}^{(r)}_{0} =& {\textstyle{1\over 4}}  \Bigl[
(I_{N}^{2} + R_{N}^{2} + I_{D}^{2} + R_{D}^{2} ) (V_4 O_4 - C_4 C_4 )+
 (I_{N}^{2} - R_{N}^{2} + I_{D}^{2} - R_{D}^{2} ) (O_4 V_4 - S_4 S_4)+
\cr
&+ 2^{r/2} (I_N I_D + R_N R_D ) (V_4 C_4 - C_4 O_4 ) + 2^{r/2} (I_N
I_D - R_N R_D ) (O_4 S_4 - S_4 V_4 ) \Bigr]
\cr}
$$
for the annulus amplitude, and
$$
\eqalign{
{\cal M}^{(r)}_{0} = &-{\textstyle{1\over 4}}  \Bigl[
(\gamma_0 + 1) I_N (\hat V _4 \hat O _4 - \hat C _4 \hat C_4 ) -
(\tilde\gamma_0 + 1) I_D (\hat V_4 \hat O_4 + \hat C_4 \hat C_4 ) +
\cr
&+ (\gamma_0 -1)  I_N (\hat O _4 \hat V _4 - \hat S_4 \hat S_4 ) -
(\tilde\gamma_0 - 1) I_D (\hat O_4 \hat V_4 + \hat S_4 \hat S_4 )  \Bigr]
\cr}
$$
for the M{\"o}bius amplitude. It is then evident that the choice $\gamma_0
= \tilde\gamma_0 = +1$ corresponds to
$$
G_{{\rm CP}} = \left. {\rm SO} (2^{4-r/2}) \otimes  {\rm SO}
(2^{4-r/2}) \right|_{99} \otimes \left. {\rm Sp} (2^{4-r/2}) \otimes  {\rm Sp}
(2^{4-r/2}) \right|_{\bar 5\bar 5}
$$
while the choice $\gamma_0 = \tilde\gamma_0 = -1$ results in unitary
gauge groups both for the D9 and D$\bar 5$-branes:
$$
G_{{\rm CP}} = \left. {\rm U} (2^{4-r/2}) \right|_{99} \otimes
\left. {\rm U} (2^{4-r/2}) \right|_{\bar 5\bar 5} \,.
$$

As pointed out in \ads , the gaugini of the unitary group are not lifted
in mass, since they are not affected by the M{\"o}bius projection.
In the following tables we report the massless spectra for the two cases.

\vbox{
\vskip 18pt
\settabs 8\columns
\hrule
\vskip 4pt
\+ 
& $r=2,$ & $n_{T}^{c}=13,$ & $n_{H}^{c}=8,$ & $G_{{\rm CP}}={\rm
SO}(8)^{2}_{99} \otimes {\rm Sp}(8)^{2}_{\bar 5\bar 5}$
\cr
\vskip 4pt
\hrule
\vskip 4pt
\+
$\qquad \phi:$ & $4\,(8,8;1,1)\oplus 4\,(1,1;8,8) \oplus 4\, (8,1;1,8)
\oplus 4\,(1,8;8,1)$
\cr
\+ 
$\qquad \psi_\alpha:$ & $(28,1;1,1)\oplus (1,28;1,1)
\oplus (1,1;28,1)\oplus (1,1;1,28)
\oplus (8,1;8,1)\oplus (1,8;1,8)$
\cr
\+ 
$\qquad \bar\psi_{\dot\alpha}:$ & $(8,8;1,1)\oplus (1,1;8,8)$
\cr
\cleartabs
\vskip 4pt
\hrule
\vskip 12pt
{\ninepoint
{\bf Table 4.} Massless spectrum for $\gamma_0 =
\tilde\gamma _0 =+1$. $\phi,\ \psi_\alpha$ and $\bar\psi_{\dot\alpha}$
denote scalar fields, left-handed spinors and right-handed spinors,
respectively.}
\vskip 18pt
}

\vbox{
\vskip 18pt
\settabs 8\columns
\hrule
\vskip 4pt
\+ 
& $r=4,$ & $n_{T}^{c}=11,$ & $n_{H}^{c}=10,$ & $G_{{\rm CP}}={\rm
SO}(4)^{2}_{99} \otimes {\rm Sp}(4)^{2}_{\bar 5\bar 5}$
\cr
\vskip 4pt
\hrule
\vskip 4pt
\+
$\qquad \phi:$ & $4\,(4,4;1,1)\oplus 4\,(1,1;4,4) \oplus 8\, (4,1;1,4)
\oplus 8\,(1,4;4,1)$
\cr
\+ 
$\qquad \psi_\alpha:$ & $(6,1;1,1)\oplus (1,6;1,1)
\oplus (1,1;6,1)\oplus (1,1;1,6)
\oplus 2\,(4,1;4,1)\oplus 2\,(1,4;1,4)$
\cr
\+ 
$\qquad \bar\psi_{\dot\alpha}:$ & $(4,4;1,1)\oplus (1,1;4,4)$
\cr
\cleartabs
\vskip 4pt
\hrule
\vskip 12pt
{\ninepoint
{\bf Table 5.} Massless spectrum for $\gamma_0 =
\tilde\gamma _0 =+1$. $\phi,\ \psi_\alpha$ and $\bar\psi_{\dot\alpha}$
denote scalar fields, left-handed spinors and right-handed spinors,
respectively.}
\vskip 18pt
}

\vbox{
\vskip 18pt
\settabs 8\columns
\hrule
\vskip 4pt
\+ 
& $r=2,$ & $n_{T}^{c}=13,$ & $n_{H}^{c}=8,$ & $G_{{\rm CP}}={\rm
U}(8)^{2}_{99} \otimes {\rm U}(8)^{2}_{\bar 5\bar 5}$
\cr
\vskip 4pt
\hrule
\vskip 4pt
\+
& & $\qquad \phi:$ & $8\,(36;1)\oplus 8\,(1;28) \oplus 8\, (8,8)$
\cr
\+ 
& & $\qquad \psi_\alpha:$ & $({\rm Adj};1)\oplus (1;{\rm Adj})
\oplus 2(8;8)$
\cr
\+ 
& & $\qquad \bar\psi_{\dot\alpha}:$ & $2\,(36;1)\oplus 2\, (1;36)$
\cr
\cleartabs
\vskip 4pt
\hrule
\vskip 12pt
{\ninepoint
{\bf Table 6.} Massless spectrum for $\gamma_0 =
\tilde\gamma _0 =-1$. $\phi,\ \psi_\alpha$ and $\bar\psi_{\dot\alpha}$
denote scalar fields, left-handed spinors and right-handed spinors,
respectively.}
\vskip 18pt
}

\vbox{
\vskip 18pt
\settabs 8\columns
\hrule
\vskip 4pt
\+ 
& $r=4,$ & $n_{T}^{c}=11,$ & $n_{H}^{c}=10,$ & $G_{{\rm CP}}={\rm
U}(4)^{2}_{99} \otimes {\rm U}(4)^{2}_{\bar 5\bar 5}$
\cr
\vskip 4pt
\hrule
\vskip 4pt
\+
& & $\qquad \phi:$ & $8\,(10;1)\oplus 8\,(1;6) \oplus 16 \, (4,4)$
\cr
\+ 
& & $\qquad \psi_\alpha:$ & $({\rm Adj};1)\oplus (1;{\rm Adj})
\oplus 4\,(4;4)$
\cr
\+ 
& & $\qquad \bar\psi_{\dot\alpha}:$ & $2\,(10;1)\oplus 2\, (1;10)$
\cr
\cleartabs
\vskip 4pt
\hrule
\vskip 12pt
{\ninepoint
{\bf Table 7.} Massless spectrum for $\gamma_0 =
\tilde\gamma _0 =-1$. $\phi,\ \psi_\alpha$ and $\bar\psi_{\dot\alpha}$
denote scalar fields, left-handed spinors and right-handed spinors,
respectively.}
\vskip 18pt
}

Thus, in the open sector supersymmetry is broken explicitly, as a
result of different representations assigned to the bosonic and
fermionic degrees of freedom in the $\bar 5 \bar 5$ sector and from a
flip of chirality of spinors in the mixed $9\bar 5$ sector.

One can show that tadpole conditions guarantee the
cancellation of the irreducible gauge and gravitational 
anomalies in $D=6$, while the residual anomaly polynomials
$$
\eqalign{
{\cal I}_8^{(r)} =& {2^{r/2} \over 64} \left( 2^{(2-r)/2} \,
{\rm tr} R^2 - {\rm tr} F_{1}^{2} - {\rm tr} F_{2}^{2} - {\rm tr}
F_{3}^{2} - {\rm tr} F_{4}^{2} \right)^2 +
\cr
&- {2^{r/2} \over 64} \left( {\rm tr} F_{1}^{2} + {\rm tr} F_{2}^{2} 
- {\rm tr} F_{3}^{2} - {\rm tr} F_{4}^{2} \right)^2 +
\cr
&+ {4+2^{r/2}\over 64} \left( {\rm tr} F_{1}^{2} - {\rm tr} F_{2}^{2} 
+ {\rm tr} F_{3}^{2} - {\rm tr} F_{4}^{2} \right)^2 +
\cr
&+ {4-2^{r/2} \over 64} \left( {\rm tr} F_{1}^{2} - {\rm tr} F_{2}^{2} 
- {\rm tr} F_{3}^{2} + {\rm tr} F_{4}^{2} \right)^2
\cr}
$$
and 
$$
\eqalign{
{\cal I}_8^{\prime\, (r)} =& {2^{r/2}\over 16} \left( 2^{-r/2}\, 
{\rm tr} R^2 - {\rm tr} F_{1}^{2} - {\rm tr} F_{2}^{2}\right)^2 
+ {2^{r/2}\over 16} \left( {\rm tr} F_{1}^{2} - {\rm tr}
F_{2}^{2}\right)^2 +
\cr
& + {2^{r/2} \over 6} {\rm tr} F_1 \left( - {\textstyle{1\over 16}}
{\rm tr} R^2 \, {\rm tr} F_2 + {\rm tr} F_{2}^{3} \right) +
{2^{r/2} \over 6} {\rm tr} F_2 \left( - {\textstyle{1\over 16}}
{\rm tr} R^2 \, {\rm tr} F_1 + {\rm tr} F_{1}^{3} \right)
\cr}
$$
do not factorise and require a generalised Green-Schwarz
mechanism \ggsm .

%%%%%%%%%%%%%%%%%%%%%%%%%%%%%%%%%%%%%%%%%%%%%%%%%%%%%%%%%%%%%%%%%%%%%%%%%

\newsec{Conclusions}

In this paper we have investigated in some detail the open descendants of the
$T^4/Z_2$ and $T^6/Z_3$ orbifolds in the presence of a non-vanishing
background for the NS-NS antisymmetric tensor, thus generalising the
constructions of \toroidal\ and of \refs{\ps ,\bs ,\twist,\gp,\chiral }. As
expected, the total size of the Chan-Paton gauge group is reduced
according to the rank of $B_{ab}$ both on the D9 and on the D5
branes. As already noticed in \refs{\witten ,\kst}, the behaviour of the
fixed points under the world-sheet parity $\Omega$ depends on the
antisymmetric tensor. As in \bs, this yields varying numbers of tensor
multiplets for the $Z_2$ case. 
This is crucial for the correct interpretation of the
amplitudes and is consistent with the constraints of two-dimensional
conformal field theory in the presence of boundaries and/or crosscaps
\fps . Moreover, as in \toroidal,
the M{\"o}bius amplitude involves sign ambiguites that
are crucial for the correct interpretation of the amplitudes and, in
the $Z_2$ orbifold,
allow one to connect continously orthogonal and symplectic gauge groups. 
These signs are needed to enforce a correct
normalisation of the M{\"o}bius amplitude, and at rational points they
reduce to the discrete Wilson lines of \twist, that affect the $P$
transformation for ${\cal M}$. We have also shown how, in this case,
couplings between twisted scalars and vector fields indeed arise,
as demanded by the generalised Green-Schwarz mechanism \ggsm . It
would be interesting to extend this construction to other six and 
four-dimensional orbifolds and to study the effect of discrete Wilson 
lines also in these cases.

We have also applied these results to a recently proposed type I
scenario, where supersymmetry is broken on the branes but is
unbroken in the bulk \ads. The outcome is, again, varying numbers of tensor
multiplets and gauge groups of reduced rank. In this case, the signs
in the M{\"o}bius amplitude allow a continous deformation from
products of orthogonal and symplectic groups to products of
unitary groups.

%\vskip 36pt

%\noindent
%{\bf Note added}. When this work was completed, a paper discussing
%related issues appeared \bst . 

\vskip 36pt 

\noindent
{\bf Acknowledgments} 
It is a pleasure to thank M. Bianchi, K. F{\"o}rger, G. Pradisi, C. Schweigert,
Ya.S. Stanev  and in particular I. Antoniadis, C. Bachas, E. Dudas and 
A. Sagnotti for very stimulating and illuminating discussions.
This work was supported in part by EEC TMR contract 
ERBFMRX-CT96-0090.

%%%%%%%%%%%%%%%%%%%%%%%%%%%%%%%%%%%%%%%%%%%%%%%%%%%%%%%%%%%%%%%%%%%%%%%%%

\listrefs

\end